\documentclass[aps,jcp,preprint,showpacs,groupedaddress]{revtex4-1}

\usepackage[intlimits]{amsmath}
\usepackage{amsfonts}
\usepackage{amssymb}
\usepackage{graphicx}
\usepackage{bm}
\usepackage{natbib}
\usepackage{amsmath}
\usepackage{wrapfig}
\usepackage{subfigure}

\usepackage{nicefrac}
\usepackage{setspace} 
\usepackage{booktabs}
\usepackage{caption}

\begin{document}

\title{Protein search for multiple targets on DNA}

\author{Martin Lange$^{1,2}$}
\author{Maria Kochugaeva$^{2}$}
\author{Anatoly B. Kolomeisky $^{2,3}$}
 \email{ tolya@rice.edu}
\affiliation{$^1$Johannes Gutenberg University, Mainz, 55122,  Germany}
\affiliation{$^2$Department of Chemistry, Rice University, Houston, Texas, 77005, USA}
\affiliation{$^3$Center for Theoretical Biological Physics, Rice University, Houston, Texas, 77005, USA}

\date{\today}

\begin{abstract}
Protein-DNA interactions are  crucial for all biological processes. One of the most important fundamental aspects of these interactions is the process of protein searching and recognizing specific binding sites on DNA.  A large number of experimental and theoretical investigations have been devoted to uncovering the molecular description of these phenomena, but many aspects of the mechanisms of protein search for the targets on DNA  remain not well understood. One of the most intriguing problems is the role of multiple targets  in protein search dynamics. Using a recently developed  theoretical framework  we analyze this question in detail. Our method is based on a discrete-state stochastic approach that takes into account  most relevant physical-chemical processes and  leads to fully analytical description of all dynamic properties. Specifically, systems with two and three targets have been explicitly investigated. It is found that multiple targets in most cases accelerate the search in comparison with a single target situation. However, the  acceleration is not always proportional to the number of targets. Surprisingly, there are even situations when it takes longer to find one of the multiple targets in comparison with the single target. It depends on the spatial position of the targets, distances between them, average scanning lengths of protein molecules on DNA, and the total DNA lengths. Physical-chemical explanations of observed results are presented. Our predictions are compared with experimental observations as well as with results from a continuum theory for the protein search. Extensive Monte Carlo computer simulations fully support our theoretical calculations.

\end{abstract}

\maketitle

\section{INTRODUCTION}

Major cellular activities are effectively governed by multiple protein-DNA interactions \cite{alberts,lodish,phillips}. The starting point of these interactions is a process of protein searching and recognizing for the specific binding sites on DNA. This is a critically important step because it allows a genetic information contained in DNA to be effectively transferred by initiating various biological processes  \cite{alberts,lodish,phillips}.  In recent years, the fundamental processes associated with the protein search for targets on DNA have been  studied extensively using a wide variety of experimental and theoretical methods \cite{riggs70,berg81,berg85,winter81,halford04,gowers05,iwahara06,kolesov07,wang06,elf07,tafvizi08,vandenbroek08,mirny09,rau10,larson11,tafvizi11,hammar12,leith12,zandarashvili12,kolomeisky12,kolomeisky11,esadze14,bauer13,brackley12,marcovitz13,landry13,koslover11,sheinman12,veksler13,sidorova13,ivanov14}.  Although a  significant progress in our understanding of the protein search phenomena has been achieved, the full description  of the mechanisms remains a controversial and highly-debated research topic \cite{mirny09,kolomeisky11,bauer13,koslover11,sheinman12,veksler13}.   

Experimental investigations of the protein search phenomena  revealed that many proteins find their  targets on DNA very fast, and the corresponding association rates might exceed the estimates from 3D diffusion limits \cite{riggs70,halford04,mirny09,kolomeisky11}. These surprising phenomena are known as a {\it facilitated diffusion} in the protein search field. More recent single-molecule experiments, which can directly visualize the dynamics of individual molecules, also suggest that during the search proteins move not only through the bulk solution via 3D diffusion but they also bind {\it non-specifically} to DNA  where they hop in 1D fashion \cite{halford04,gowers05,wang06,elf07,tafvizi11,hammar12,landry13}.  Several theoretical approaches that incorporate the coupling between 3D diffusion and 1D sliding in the protein search have been proposed \cite{halford04,mirny09,bauer13,kolomeisky11,veksler13}, but they had a variable success in explaining all experimental observations. 

One of the most interesting problems related to the protein search on DNA is the effect of multiple targets. The question is how long will it take for the proteins to find {\it any} specific binding site from several targets present on DNA.  Naively, one could argue that in this case the search time should be accelerated proportionally to the number of targets, i.e., the association reaction rate  should be proportional to the concentration of specific binding sites. However, this effectively mean-field view  ignores several important observations. First, it is clear that the search time for several targets lying very close to each other generally should not be the same as the search time for the same number of targets which are spatially dispersed. Second, the experimentally supported complex 3D+1D search mechanism  suggests that varying spatial distributions of the specific binding sites should also affect the search dynamics. Thus, it seems that the simple mean-field arguments should not be valid for all conditions. Surprisingly, this very important problem was addressed only in one recent work \cite{hammar12}. Hammar et al. \cite{hammar12} using high-quality single-molecule measurements in the living cells investigated the dynamics of finding the specific sites for  {\it lac} repressor proteins on DNA with two targets. It was found that the association rates increase as a function of the distance between targets \cite{hammar12}. An approximate theoretical model for the protein search with two targets was proposed. However, this theoretical approach has several problems. It was presented for infinitely long DNA chains using a continuum approximation. At the same time, it was shown recently that the continuum approach might lead to serious errors and artifacts in the description of protein search dynamics \cite{veksler13}.  In addition, this theory predicted that the acceleration due to the presence of two targets in comparison with the case of only one target should disappear in the limit of very large sliding lengths. This is clearly a nonphysical result. In this limit, the protein spends most of the searching time on DNA and it is faster to find any of two targets than one specific binding site. 

In this article, we present a comprehensive theoretical method of analyzing the role of multiple targets in the protein search on DNA. Our approach is based on a discrete-state stochastic framework that was recently developed by one of us  for the search with one specific binding site \cite{veksler13}. It takes into account most relevant biochemical and biophysical processes, and it allows us to obtain fully analytical solutions for all dynamic properties at all conditions. One of the main results of the  discrete-state stochastic method was a construction of dynamic phase diagram  \cite{veksler13}. Three possible dynamic search regimes were identified. When the protein sliding length was larger than the DNA chain length, the search followed  simple random-walk dynamics with a quadratic scaling of the search time on the DNA length. For the sliding length smaller than the DNA length but larger than the the size of the specific binding site, the search dynamic followed a linear scaling. When the sliding length was smaller than the target size, the search was dominated by nonspecific bindings and unbindings  without the sliding along  DNA. In this paper, we extend this method to the case of several specific binding sites at arbitrary spatial positions. It allows us to explicitly describe the role of multiple targets and their spatial distributions in the protein search. Our theoretical calculations agree with available experimental observations, and we also test them in  Monte Carlo computer simulations.

\section{THEORETICAL METHOD} 
 
The original discrete-state stochastic approach can be generalized for any number of the specific binding sites at arbitrary positions along the DNA chain. But to explain the main features of our theoretical method, we analyze specifically a simpler model with only two targets as shown in Fig. 1. A single DNA molecule with $L$ binding sites and a single protein molecule  are considered. The analysis can be easily extended for any concentration of proteins and DNA \cite{esadze14}. Two of the bindings sites $i$ and $j$ ($i=m_{1}$ and $j=m_{2}$) are targets for the protein search (see Fig. 1). The protein starts from the bulk solution that we label as a state $0$. Since 3D diffusion is usually much faster than other processes in the system, we assume that the protein can access with equal probability any site on the DNA chain (with the corresponding total binding rate $k_{on}$). While being on DNA, the protein can move with a diffusion rate $u$ along the chain with equal probability in both directions. The protein molecule can also dissociate from DNA with a rate $k_{off}$ to the bulk solution (Fig. 1). The search process ends when the protein reaches for the first time {\it any} of two targets. 

The main idea of our approach is to utilize first-passage processes to describe the complex dynamics of the protein search on DNA \cite{veksler13}. One can introduce a function $F_{n}(t)$ defined as a probability to reach any target at time $t$ for the first time if initially (at $t=0$) the protein molecule starts at the state $n$ ($n=0,1,\ldots,L$). These first-passage probabilities evolve with time as described by a set of the backward master equations \cite{kolomeisky12,veksler13},
\begin{equation}\label{me1}
\frac{dF_n(t)}{dt}=u [F_{n+1}(t)+F_{n-1}(t)]+k_{off} F_0(t)-(2u+k_{off}) F_n(t),
\end{equation}
for $2 \le n \le L-1$. At DNA ends ($n=1$ and $n=L$) the dynamics is slightly different,
\begin{equation}\label{me2}
\frac{dF_1(t)}{dt}=u F_2(t)+k_{off} F_0(t)-(u+k_{off}) F_1(t);
\end{equation}
and
\begin{equation}\label{me3}
\frac{F_L(t)}{dt}=u F_{L-1}(t)+k_{off} F_0(t)-(u+k_{off}) F_L(t).
\end{equation}
In addition, in the bulk solution we have
\begin{equation}\label{me4}
\frac{dF_0(t)}{dt}=\frac{k_{on}}{L} \sum_{n=1}^{L} F_n(t) - k_{on} F_0(t).
\end{equation}
Furthermore, the initial conditions require that
\begin{equation}
F_{m_{1}}(t)=F_{m_{2}}(t)=\delta(t), \quad F_{n\ne m_{1},m_{2}}(t=0)=0.
\end{equation}
The physical meaning of this statement is that if we start at one of two targets the search process is finished immediately.

It is convenient to solve Eqs. (\ref{me1}), (\ref{me2}), (\ref{me3}) and (\ref{me4}) by employing Laplace transformations of the first-passage probability functions, $\widetilde{F_{n}(s)} \equiv \int_0^{\infty}e^{-st}F_n(t)dt$ \cite{veksler13}. The details of calculations are given in the Appendix. It is important to note here that the explicit expressions for the first-passage probability distribution functions in the Laplace form  provide us with a full dynamic description of the protein search \cite{veksler13}. For example, the mean first-passage time to reach any of the target sites if the original position of the protein was in the solution ($n=0$), which we also associate with the search time, can be directly calculated from \cite{veksler13}
\begin{equation}
T_{0} \equiv -\frac{\partial \widetilde{F_{0}(s)}}{\partial s}|_{s=0}.
\end{equation}
As shown in the Appendix, the average search time is given by
\begin{equation}\label{eqT0}
T_0 = \frac{k_{off} L + k_{on} [L-S_{i}(0)]}{k_{on} k_{off} S_{i}(0)},
\end{equation}
where $S_{i}(s)$ is a new auxiliary function with a subscript specifying the number of targets ($i=2$ for the system with two targets). For this function we have
\begin{equation}
S_2(s) = \frac{(1+y)\left[2(1-y^{2 L+m_1-m_2})+(1-y^{m_2-m_1}) (y^{2 m_1-1}+y^{1+2(L-m_2)})\right]}{(1-y)(1+y^{2 m_1 -1})(1+y^{1+2(L-m_2)})(1+y^{m_2-m_1})},
\end{equation}
with
\begin{equation}
y = \frac{s + 2 u + k_{off} - \sqrt{(s + 2 u + k_{off})^2 - 4 u^2}}{2 u}.
\end{equation}
It is important that for $m_{1}=m_{2}$, as expected, our results reduce to expressions for the protein search on DNA with only one target \cite{veksler13}. Similar procedures can be used to estimate all other dynamic properties for the system with two targets. 

We can extend this approach for any number of targets and for any spatial distribution of binding sites. This is discussed in detail in the Appendix. Surprisingly, the expression for the search times are the {\it same} in all cases but with different $S_{i}$ functions that depend on the number of specific binding sites and their spatial distributions. Analytical results for $S_{i}$ for the protein search on DNA with  three or four targets, as well as a general procedure for arbitrary number of specific binding sites, are also presented in the Appendix.

\section{RESULTS AND DISCUSSION}

\subsection{Spatial Distribution of Targets}

Because our theoretical method provides explicit formulas for all relevant quantities, it allows us to fully explore many aspects of the protein search mechanisms. The first problem that we can address is related to the role of the spatial distribution of targets on the search dynamics. In other words, the question is how the search time is influenced by exact positions of all targets along the DNA. The results of our calculations for two specific binding sites are presented in Fig. 2. The longest search times are found when two targets are at different ends of the molecule, and the  distance between them along the DNA curve, $l=|m_{1}-m_{2}|$, is the largest  possible and equal to $L-1$. The search is faster if targets are  moved closer to each other and both distributed symmetrically with respect to the middle point of the DNA molecule (Fig. 2).  Moving the targets too close ($ l \simeq 0$) starts to increase the search time again: see Fig. 2. For short DNA chains, it can be shown that there is an optimal distance between two targets, $l_{opt}=L/2$, that yields the fastest search (Fig. 2). It corresponds to  the most optimal positions of the specific sites to be at $m_{1}=L/4$ and $m_{2}=3L/4$.

The last result is slightly unexpected since simple  symmetry arguments suggest that the fastest search  would be observed for the uniform distribution of targets, i.e. when the distance between the specific sites and the distance between the ends and targets are the same, i.e., for $m_{1}=L/3$ and $m_{2}=2L/3$. This is not  observed in Fig. 2. To explain this, one can argue that the search on the DNA molecule of length $L$ with $n$ targets can be mapped into the search on $n$ DNA segments of variable lengths  with only one target per each segment. In this case, positioning  each target in the middle of the corresponding segment leads to the fastest search dynamics.\cite{veksler13} This suggests that the most optimal distribution of $n$ symmetrically distributed targets is a uniform distribution with the distance between two neighboring targets equal to $L/n$. But then the first and the last targets will be separated from the corresponding ends by a shorter distance $\frac{L-(n-1)\frac{L}{n}}{2}=\frac{L}{2n}$. This is exactly what we see in Fig. 2 for $n=2$ targets. The reason that distances between the ends and the closest targets deviate from the distances between the targets is the reflecting boundary conditions at the ends that are assumed in our model: see Eqs. (\ref{me2}) and (\ref{me3}).

The results presented in Fig. 2 also illustrate another interesting observation. Increasing the length of DNA effectively eliminates the minimum in the search time for specific symmetric locations of the targets. Essentially, for $L \gg 1$, which is much closer to realistic conditions in most cases, any two position of the targets inside the DNA chain will be optimal and will have the same search time as long as they are not at the ends. We will discuss the reason for this below.

\subsection{Dynamic Phase Diagram}

One of the main advantages of our method is the ability to explicitly analyze the search dynamics for all ranges of relevant parameters. This allows us to construct a comprehensive dynamic phase diagram that delineates different search regimes. The results are presented in Fig. 3 for the systems with different numbers of specific binding sites. The important observation is that general features of the search behavior are independent of the number of targets. 

More specifically, there are three dynamic phases that depend on the relative values of the length of DNA $L$, the average scanning length $\lambda=\sqrt{u/k_{off}}$ and the size of the target (taken to be equal to unity in our model).  For $\lambda > L$ the random-walk regime is observed with the search time being quadraticaly proportional to the size of DNA \cite{veksler13}. In this case, the protein non-specifically binds  DNA and it does not dissociate until it finds one of the targets. The quadratic scaling is a result of a simple random-walk unbiased diffusion of the protein molecule on DNA during the search. For the intermediate sliding regime, $1 < \lambda <L$, the protein binds to DNA, scans it, unbinds and repeats this cycle at average $L/n\lambda$ times ($n$ is number of the targets) for symmetrically distributed specific sites. For more general distributions the number of search cycles is also proportional to $L/\lambda$. This leads to the linear scaling in the search times. For $\lambda <1$ we have the jumping regime where the protein can bind to any site on DNA and dissociate from it, but it cannot slide along the DNA chain. The search time is again proportional to $L$ because on average the protein must check $L$ sites. These changes in the dynamic search behavior are illustrated in Fig. 4, in which the search times as a function of the DNA lengths are presented for different scanning lengths. The slope variation indicates a change in the scaling behavior in the search times from $L^{2}$ to $L$ as the DNA length increases for fixed $\lambda$.

It is also important to note here that the concept of the most optimal positions of targets is not working for the sliding  regime ($ 1 < \lambda < L$) because the protein during the search frequently unbinds from the DNA, losing all memory about what it already scanned. This concept also cannot be defined in the jumping regime where the protein does not slide at all. From this point of view, any position of the targets are equivalent. The only two positions that differ from others are the end sites in the sliding regime. This is because they can be reached only via one neighboring site, while all other sites can be reached via two neighboring sites (see Fig. 1).

\subsection{Acceleration of the Search}

The most interesting question for this system is to analyze quantitatively the effect of multiple targets on search dynamics. To quantify this we define a new function, $a_{n}$, which we call an acceleration,
\begin{equation}
a_{n}=\frac{T_{0}(1)}{T_{0}(n)}.
\end{equation}
This is a ratio of the search times for the case of one target and for the case of $n$ targets. The parameter $a_{n}$ gives a numerical value of how the presence of multiple targets increases the rate of association to any specific binding site. The results for acceleration are presented in Figs. 5-7.

First, we analyze the situation when targets in all cases are in the most optimal symmetric positions, which is shown in Fig. 5. For DNA with the single target it is in the middle of the chain, while for DNA with $n$ targets they are distributed uniformly, as we discussed above, with the distance $L/n$ between the internal targets and $L/2n$ for boundary targets and DNA ends. The acceleration for these conditions depends on the dynamic search regimes, and it ranges from $n$ to $n^{2}$: see Fig. 5. For the case of $\lambda < L$ (jumping and sliding regimes), on average the number of search excursions to DNA before finding the specific site is equal to $L/n$, and this leads to a linear behavior in the acceleration ($a_{n} \simeq n$). For $\lambda > L$ (random-walk regime), the search is one-dimensional and the protein must diffuse on average the distance $L/n$ before it can find any of the targets. The quadratic scaling for the simple random walk naturally explains the acceleration in the search in this dynamic regime, $a_{n} \simeq n^{2}$.  

However, the acceleration is also affected by the distance between the targets. If we maintain the most optimal conditions for the DNA with one target but vary the distance between multiple targets, while keeping the overall symmetry, the results are shown in Fig. 6. In this case, putting targets too close to each other or moving them apart lowers the acceleration. Eventually, there will be no acceleration for these conditions ($a_{n}=1$). But the results are much more interesting if we consider the non-symmetric distributions of targets. Surprisingly, the search time for the system with multiple targets can be even slower than for the single target system! This is shown in Fig. 7 where $a_{n}$ can be as low as 1/4 for the two-target system in the random-walk regime, or it can reach the value of 1/2 in the sliding and jumping regimes (not shown). The single target in the most optimal position in the middle of the DNA chain can be found much faster in comparison with the case of two targets seating near one of the ends.

These observations suggest that the degree of acceleration of the search process due to the presence multiple targets is not always a linear function of the number of specific binding sites. It depends on the nature of the dynamic search phase, the distance between the targets and the spatial distribution of the targets. Varying these parameters can lead to larger accelerations as well as to unexpected decelerations. It is a consequence of the complex mechanism of the protein search for targets on DNA that combines 3D and 1D motions. This is the main result of our paper.

\subsection{Comparison with Continuum Model and with Experiments}

Recently, single-molecule experiments measured the facilitated search of {\it lac} repressor proteins on DNA  with two identical specific binding sites \cite{hammar12}. These experiments show that the association rate increases before reaching the saturation with the increase in the distance between the targets. Our theoretical model successfully describes these measurements, as shown in Fig. 8. Fitting these data, we  estimate the 1D diffusion rate for the lac repressors  as $u \simeq 7 \times 10^{5}$ s$^{-1}$, which is consistent with {\it in vitro} measured values \cite{elf07}. Our estimates for the sliding length, $\lambda \simeq 25$ bp, and for the non-specific association to DNA, $k_{on} \simeq 6.4 \times 10^{4}$ s$^{-1}$,  also agree with experimental observations \cite{hammar12}.  

It is  important to compare our results with predictions from the theoretical model presented in Ref \cite{hammar12}. This continuum model was developed assuming that the length of DNA is extremely long, $ L \gg 1$.  It was shown  that the acceleration for the search for the case of two targets can be simply written as \cite{hammar12}
\begin{equation}\label{continuum}
a_{2}=1+\tanh\left(\frac{l}{2\lambda}\right),
\end{equation} 
where $l$ is the distance between targets. The comparison between two theoretical approaches is given in Figs. 9 and 10. One can see from Fig. 9 that both models agree for very large DNA lengths, $L \gg 1$, while for  shorter DNA chains there are significant deviations. The continuum theory \cite{hammar12} predicts that the acceleration is always a linear or sub-linear function of the number of targets, i.e.,  $1 \le a_{n} \le n$. Our model shows that the acceleration can have a non-linear dependence on the number of the targets, $a_{n} \simeq n^{2}$. More specifically, this can be seen  in Fig. 10, where the acceleration is presented as a function of the scanning length $\lambda$. The prediction of the continuum theory that for $\lambda \gg 1$ the acceleration always approaches the unity is unphysical. Clearly, if we consider, e.g.,  the optimal distribution of targets, then the larger the number of specific binding sites, the shorter the search time.  The reason for the failure of the continuum model at this limit is its inability to properly account for all dynamic search regimes. This analysis shows that the continuum model \cite{hammar12} has a very limited application, while our theoretical approach is consistent with all experimental observations and provides a valid physical picture for all conditions.

\section{SUMMARY AND CONCLUSIONS}

We investigated theoretically the effect of the multiple targets in the protein search for specific binding sites on DNA. This was done by extending and generalizing the discrete-state stochastic method, originally developed for single targets, that explicitly takes into account the most important biochemical and biophysical processes. Using the first-passage processes, all dynamic properties of the system can be directly evaluated. It was found that the search dynamics is affected by the spatial distribution of the targets for not very long DNA chains. There are  optimal positions for specific sites  for which the search times are minimal. We argued that this optimal distribution is almost uniform with a correction due to the DNA chain ends. We also constructed a dynamic phase diagram for the different search regimes. It was shown that for any number of targets  there are always three phases, which are determined by comparing the DNA length, the scanning length and the size of the target. Furthermore, we investigated the quantitative acceleration in the search  due to the presence of multiple targets for various sets of conditions. It was found also that the acceleration is linearly proportional  to the number of targets when the scanning length is less than the DNA length. For larger scanning lengths, the acceleration becomes faster with the quadratic dependence on the number of targets.  However, changing the distances between the targets generally decreases the effect of acceleration. Unexpectedly, we found that varying also the spatial distributions  can reverse the behavior: it might take longer to find the specific site in the system with multiple targets in comparison with properly positioned single target. Our model allows us to explain this complex behavior using simple physical-chemical arguments. In addition, we applied our theoretical analysis for describing experimental data, and it is shown that the obtained dynamic parameters are consistent with measured experimental quantities. A comparison between our discrete-state theoretical method the continuum model is also presented. We show that the continuum model has a limited range of applicability, and it produces the unphysical behavior at some limiting cases. At the same time, our approach is fully consistent at all sets of parameters. Our theoretical predictions were also fully validated with Monte Carlo computer simulations.

The presented theoretical model seems to be successful in explaining the complex protein search dynamics in the systems with multiple targets. One of the main advantage of the method is the ability to have a fully analytical description for all dynamic properties in the system. However, one should remember that this approach is still quite oversimplified, and it neglects many realistic features of the protein-DNA interactions. For example, DNA molecule is assumed to be frozen, different protein conformations that are observed in experiments are not taken into account, and the possibility of correlations between 3D and 1D motions is also not considered. It will be critically important to test the presented theoretical ideas in experiments as well as in more advanced theoretical methods.

\begin{acknowledgments}
The work was supported by the Welch Foundation (Grant C-1559), by the NSF (Grant CHE-1360979), and by the Center for Theoretical Biological Physics sponsored by the NSF (Grant PHY-1427654).
\end{acknowledgments}

\section*{Appendix: Explicit calculations of first-passage probability functions and average search times}

This appendix includes detailed derivations of the equations from the main text and explicit expressions for functions utilized in our calculations.
 
To solve the backward master equations (1)-(5)  for the system with two targets we use the Laplace transformation which leads to
\begin{equation}\label{l1}
(s+2 u+k_{off}) \widetilde{F_n}(s)=u \left[\widetilde{F_{n+1}}(s)+\widetilde{F_{n-1}}(s)\right]+k_{off} \widetilde{F_0}(s);
\end{equation}
\begin{equation}\label{l2}
(s+u+k_{off})\widetilde{F_1}(s)=u \widetilde{F_2}(s)+k_{off} \widetilde{F_0}(s);
\end{equation}
\begin{equation}\label{l3}
(s+u+k_{off})\widetilde{F_L}(s)=u \widetilde{F_{L-1}}(s)+k_{off} \widetilde{F_0}(s);
\end{equation}
\begin{equation}\label{l4}
(s+k_{on})\widetilde{F_0}(s)=\frac{k_{on}}{L} \sum_{n=1}^{L} \widetilde{F_n}(s);
\end{equation}
with the condition that
\begin{equation}
\widetilde{F_{m_1}}(s)=\widetilde{F_{m_2}}(s) = 1.
\end{equation}

We are looking for the solution of these equations  in the form, $\widetilde{F_n}(s)=A\cdot y^{n} + B$, where $A$ and $B$ are unknown coefficients that will be determined after the substitution of the solution into Eqs. (\ref{l1}), (\ref{l2}), (\ref{l3}) and (\ref{l4}). This gives the following expression,
\begin{equation}
(s+2 u+k_{off})(A y^{n} + B)=u\left[A y^{n+1} + B + A y^{n-1} + B\right]+k_{off} \widetilde{F_0}(s).
\end{equation}
After rearranging, we obtain
\begin{equation}
A\left[u y^{n+1} - (s+2 u+k_{off}) y^n + u y^{n-1}\right] = (s + k_{off}) B - k_{off} \widetilde{F_0}(s).
\end{equation}
Requiring  that the right-hand-side of this expression  to be equal to zero, yields
\begin{equation}\label{eqB}
B = \frac{k_{off}}{s+k_{off}} \widetilde{F_0}(s)
\end{equation}
Since the parameter $A \neq 0$, we can find $y$ by solving
\begin{equation}
u y^{n+1} - (s+2 u+k_{off}) y^n + u y^{n-1} = 0,
\end{equation}
or
\begin{equation}
u y^{2} - (s+2 u+k_{off}) y + u = 0.
\end{equation}
There are two roots of this quadratic equation,
\begin{equation}
y_{1} = \frac{s + 2 u + k_{off} - \sqrt{(s + 2 u + k_{off})^2 - 4 u^2}}{2 u},
\end{equation}
and 
\begin{equation}
y_{2} = \frac{s + 2 u + k_{off} + \sqrt{(s + 2 u + k_{off})^2 - 4 u^2}}{2 u},
\end{equation}
with $y_{2}=1/y_{1}$.

The next step is to notice that two targets at the positions $m_{1}$ and $m_{2}$ divide the DNA chain into 3 segments which can be analyzed separately. Then the general solution should have the form
\begin{equation}
\widetilde{F_n}(s)=A_1 y^{n} + A_2 y^{-n} + B,
\end{equation}
with the parameter $B$ is specified by Eq. (\ref{eqB}) and $y=y_{1}$. Using the corresponding boundary conditions, it can be shown that for $1 \le n \le m_{1}$
\begin{equation}
\widetilde{F_{n}}(s) = \frac{(1-B) \left(y^n+y^{1-n}\right)}{y^{m_1}+y^{1-m_1}}+B,
\end{equation}
while for $m_{1} \le n \le m_{2}$ we have
\begin{equation}
\widetilde{F_{n}}(s)=\frac{(1-B) \left(y^n+y^{m_1+m_2-n}\right)}{y^{m_1}+y^{m_2}}+B,
\end{equation}
and for $m_{2} \le n \le L$
\begin{equation}
\widetilde{F_{n}}(s)=\frac{(1-B)\left(y^{n-L}+y^{1+L-n}\right)}{y^{m_2-L}+y^{1+L-m_2}} + B
\end{equation}
This leads to the following expression for $\widetilde{F_0}(s)$:
\begin{equation}\label{eqF0}
\widetilde{F_0}(s)=\frac{k_{on} (k_{off}+s)  S_{i}(s)}{L  s (s+k_{on}+k_{off})+k_{on} k_{off} S_{i}(s)},
\end{equation}
where the auxiliary function $S_{i}(s)$ is introduced via the following relation
\begin{equation}
\sum_{n=1}^{L} \widetilde{F_n}(s)=(1-B)S_{i}(s)+BL.
\end{equation}
Note that Eq.(\ref{eqF0}) is identical to the corresponding equation for the single-target case \cite{veksler13}, but with the different auxiliary function $S_{i}(s)$.

Finally, we can obtain the explicit expressions for the search times as given in the main text in Eq. (7). The explicit form of the search time depends on the auxiliary functions $S_{i}$, which can be directly evaluated. For example, for the two targets we have
\begin{equation}
S_{2}(s) = \sum_{n=1}^{m_1-1} \frac{y^n+y^{1-n}}{y^{m_1}+y^{1-m_1}} + \sum_{n=m_1}^{m_2-1} \frac{y^n+y^{m_1+m_2-n}}{y^{m_1}+y^{m_2}} + \sum_{n=m_2}^L \frac{y^{n-L}+y^{1+L-n}}{y^{m_2-L}+y^{1+L-m_2}},
\end{equation}
which after simplifications leads to Eq. (8) in the main text. Similar analysis can be done for any number of targets with arbitrary positions along the chain. The final expression for the search times is the same in all cases [given by the Eq.(7)], but with the different auxiliary functions $S_{i}(s)$.  When the protein molecule searches the DNA with three targets ($i=3$), it can be shown that
\renewcommand{\arraystretch}{2}
\begin{equation}
S_{3}(s)=\frac{1}{y-1}\left[\frac{y^{2+2L} - y^{2m_3}}{y^{1+2L} + y^{2m_3}} - \frac{y^2 - y^{2m_1}}{y + y^{2m_1}} - (1+y)\left(\frac{y^{m_1} - y^{m_2}}{y^{m_1} + y^{m_2}} + \frac{y^{m_2} - y^{m_3}}{y^{m_2} + y^{m_3}}\right)\right].
\end{equation}
For the system with four targets ($i=4$) we obtain
\begin{equation}
S_4(s)=\frac{1}{y-1}\left[\frac{y^{2+2L} - y^{2m_4}}{y^{1+2L} + y^{2m_4}} - \frac{y^2 - y^{2m_1}}{y + y^{2m_1}} - (1+y)\left(\frac{y^{m_1} - y^{m_2}}{y^{m_1} + y^{m_2}} + \frac{y^{m_2} - y^{m_3}}{y^{m_2} + y^{m_3}} + \frac{y^{m_3} - y^{m_4}}{y^{m_3} + y^{m_4}}\right)\right].
\end{equation}

\newpage

\noindent Fig. 1.  A general view of  the discrete-state stochastic model for the protein search on DNA with two targets. There are $L-2$ nonspecific and $2$ specific binding sites on the DNA chain. A protein molecules can diffuse along the DNA with the rate $u$, or it might dissociate into the solution with the rate $k_{off}$. From the solution, the protein can attach to any position on DNA with the total rate $k_{on}$. The search process is considered to be completed when the protein binds for the first time to any of two targets at the position $m_{1}$ or $m_{2}$. 

\vspace{5mm}

\noindent Fig. 2. Normalized search times as a function of the normalized distance between two targets. The targets are positioned symmetrically with respect to the center of the DNA chain. The parameters used in calculations are the following: $u=k_{on}=10^{5}$ s$^{-1}$ and $k_{off}=10^{3}$ s$^{-1}$. The scanning length $\lambda$ is varied by changing $k_{off}$.  Solid curves are theoretical predictions, symbols are from Monte Carlo computer simulations.

\vspace{5mm}

\noindent Fig. 3. Dynamic phase diagram for the protein search with multiple targets. Search times as a  function of the scanning length are shown for systems with one, two or three targets. The parameters used in calculations are the following: $L=10001$ bp; and $u=k_{on}=10^{5}$ s$^{-1}$. The scanning length $\lambda$ is varied by changing $k_{off}$.

\vspace{5mm}

\noindent Fig. 4. Protein search times as a function of DNA length for different scanning lengths for the system with two targets.  The parameters used in calculations are the following: $u=k_{on}=10^{5}$ s$^{-1}$. Solid curves are theoretical predictions, symbols are from Monte Carlo computer simulations. The scanning length $\lambda$ is varied by changing $k_{off}$.

\vspace{5mm}

\noindent Fig. 5. Acceleration in the search times as a function of the scanning length for the systems with two and three targets. The parameters used in calculations are the following: $u=k_{on}=10^{5}$ s$^{-1}$. The scanning length $\lambda$ is varied by changing $k_{off}$.

\vspace{5mm}

\noindent Fig. 6. Acceleration in the search times as a function of the normalized distance between the targets for the systems with two and three targets. The single target is in the middle of the DNA chain. Other targets systems are symmetric but not optimal. The parameters used in calculations are the following: $u=k_{on}=10^{6}$ s$^{-1}$; $k_{off}=10^{-4}$ s$^{-1}$ and $L=10^{5}$ bp. 

\vspace{5mm}

\noindent Fig. 7. Acceleration in the search times as a function of the normalized distance between the targets for the systems with  targets. The single target is in the middle of the DNA chain. In the two-target system one of the specific binding sites is fixed at the end and the position of the second one is varied.  The parameters used in calculations are the following: $u=k_{on}=10^{5}$ s$^{-1}$; $k_{off}=10^{-4}$ s$^{-1}$ and $L=10^{5}$.

\vspace{5mm}

\noindent Fig. 8. Describing the experimental data from Ref. \cite{hammar12} using Eq. (7). Parameters obtained from the best fit are discussed in the text.

\vspace{5mm}

\noindent Fig. 9. Comparison of theoretical predictions for the acceleration as a function of the distance between the specific binding sites  for the system with two targets for different DNA lengths.  Targets are distributed symmetrically with respect to the middle of the DNA chain. Solid curves are discrete-state predictions, dashed curves are from the continuum model from Ref. \cite{hammar12}. The parameters used in calculations are the following: $u=k_{on}=10^{5}$ s$^{-1}$; and $k_{off}=10^{3}$ s$^{-1}$.

\vspace{5mm}

\noindent Fig. 10. Comparison of theoretical predictions for the acceleration as a function of the scanning length  for the system with two targets for different DNA lengths. Targets are in the most optimal symmetric positions. Solid curves are discrete-state predictions, dashed curves are from the continuum model from Ref. \cite{hammar12}. The parameters used in calculations are the following: $u=k_{on}=10^{5}$ s$^{-1}$. The scanning length $\lambda$ is varied by changing $k_{off}$.

\newpage

\begin{figure}[h]
  \begin{center}
    \unitlength 1in  
    \resizebox{3.375in}{2.5in}{\includegraphics{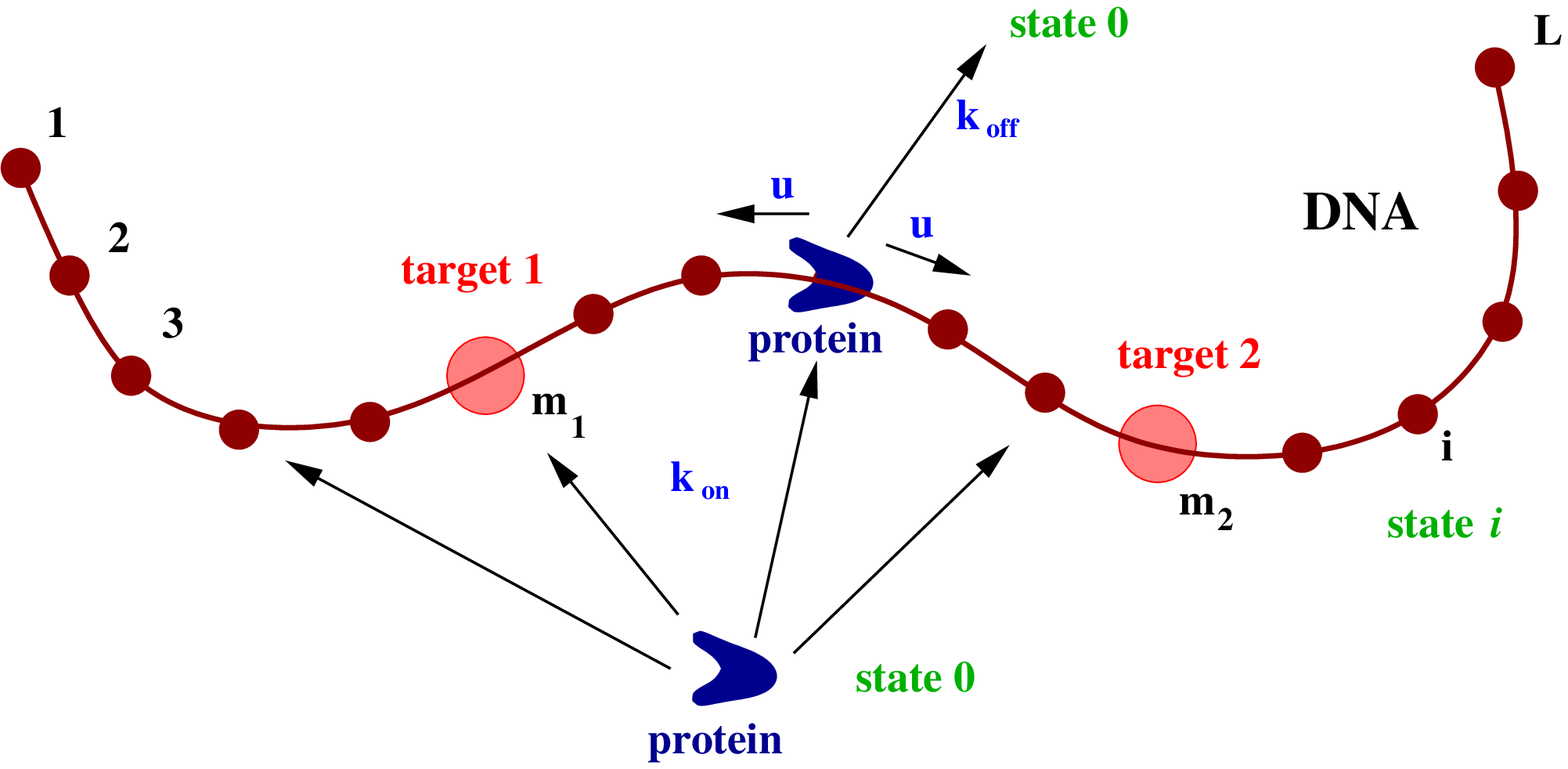}}
    \vskip 1in
    \begin{Large} Figure 1. Lange, Kochugaeva and Kolomeisky \end{Large}
    \caption{}
    \label{fig:1}
  \end{center}
\end{figure}

\newpage

\begin{figure}[h]
  \begin{center}
    \unitlength 1in  
    \resizebox{3.375in}{3.0in}{\includegraphics{Fig2.eps}}
    \vskip 1in
    \begin{Large} Figure 2. Lange, Kochugaeva and Kolomeisky \end{Large}
    \caption{}
    \label{fig:2}
  \end{center}
\end{figure}

\newpage

\begin{figure}[h]
  \begin{center}
    \unitlength 1in  
    \resizebox{3.375in}{3.0in}{\includegraphics{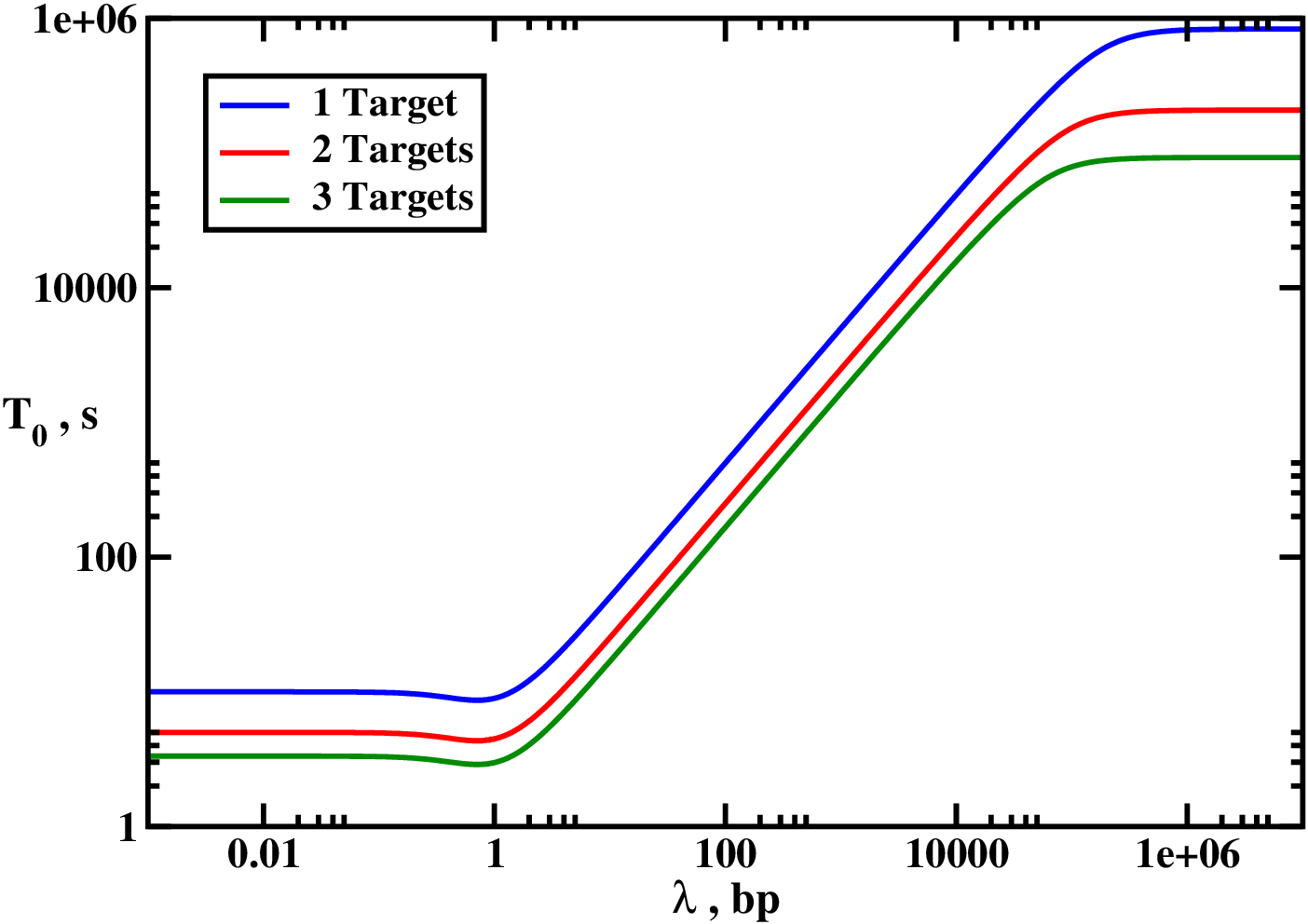}}
    \vskip 1in
    \begin{Large} Figure 3. Lange, Kochugaeva and Kolomeisky \end{Large}
    \caption{}
    \label{fig:3}
  \end{center}
\end{figure}

\newpage

\begin{figure}[h]
  \begin{center}
    \unitlength 1in  
    \resizebox{3.375in}{3.0in}{\includegraphics{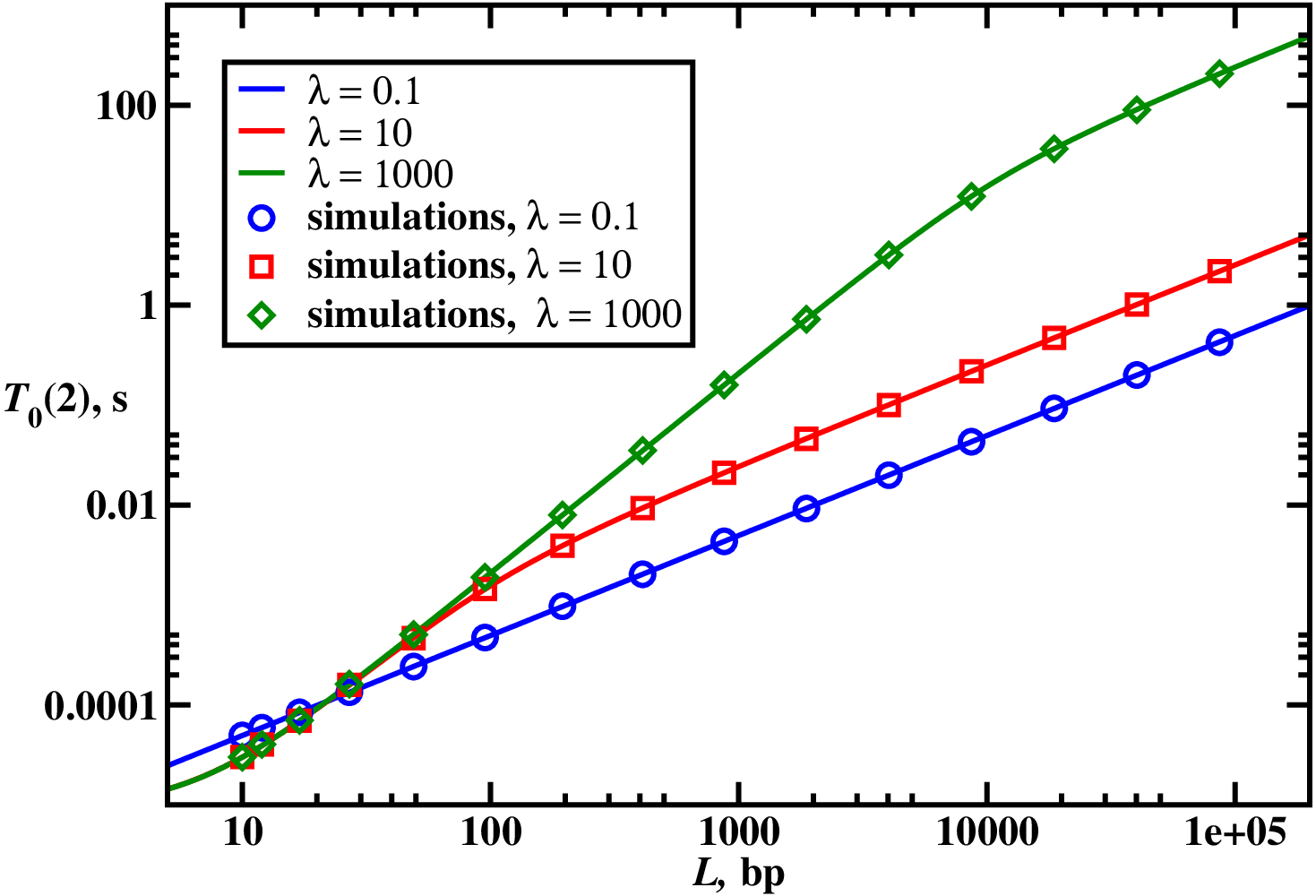}}
    \vskip 1in
    \begin{Large} Figure 4. Lange, Kochugaeva and Kolomeisky \end{Large}
    \caption{}
    \label{fig:4}
  \end{center}
\end{figure}

\newpage

\begin{figure}[h]
  \begin{center}
    \unitlength 1in  
    \resizebox{3.375in}{3.0in}{\includegraphics{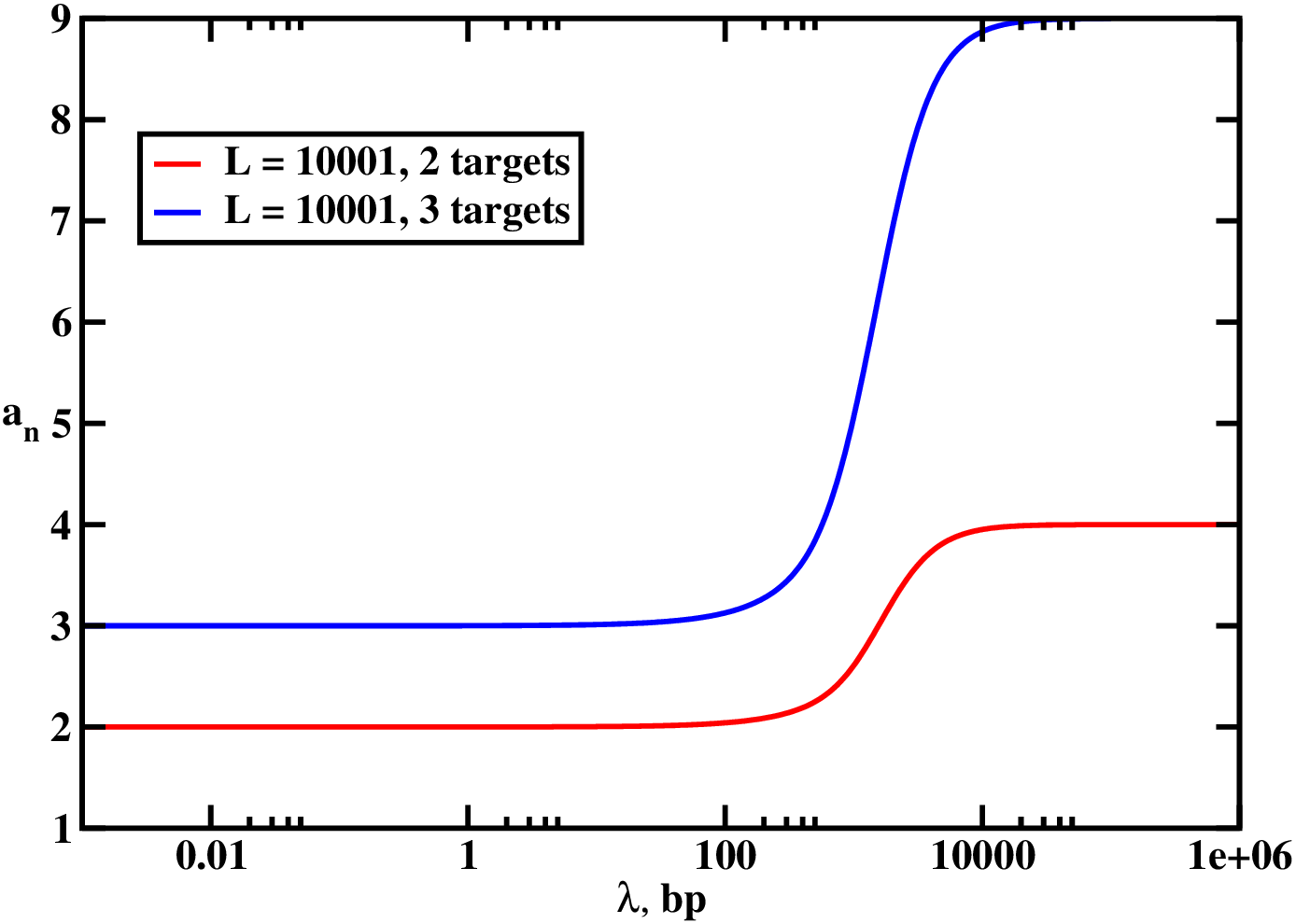}}
    \vskip 1in
    \begin{Large} Figure 5. Lange, Kochugaeva and Kolomeisky \end{Large}
    \caption{}
    \label{fig:5}
  \end{center}
\end{figure}

\newpage

\begin{figure}[h]
  \begin{center}
    \unitlength 1in  
    \resizebox{3.375in}{3.0in}{\includegraphics{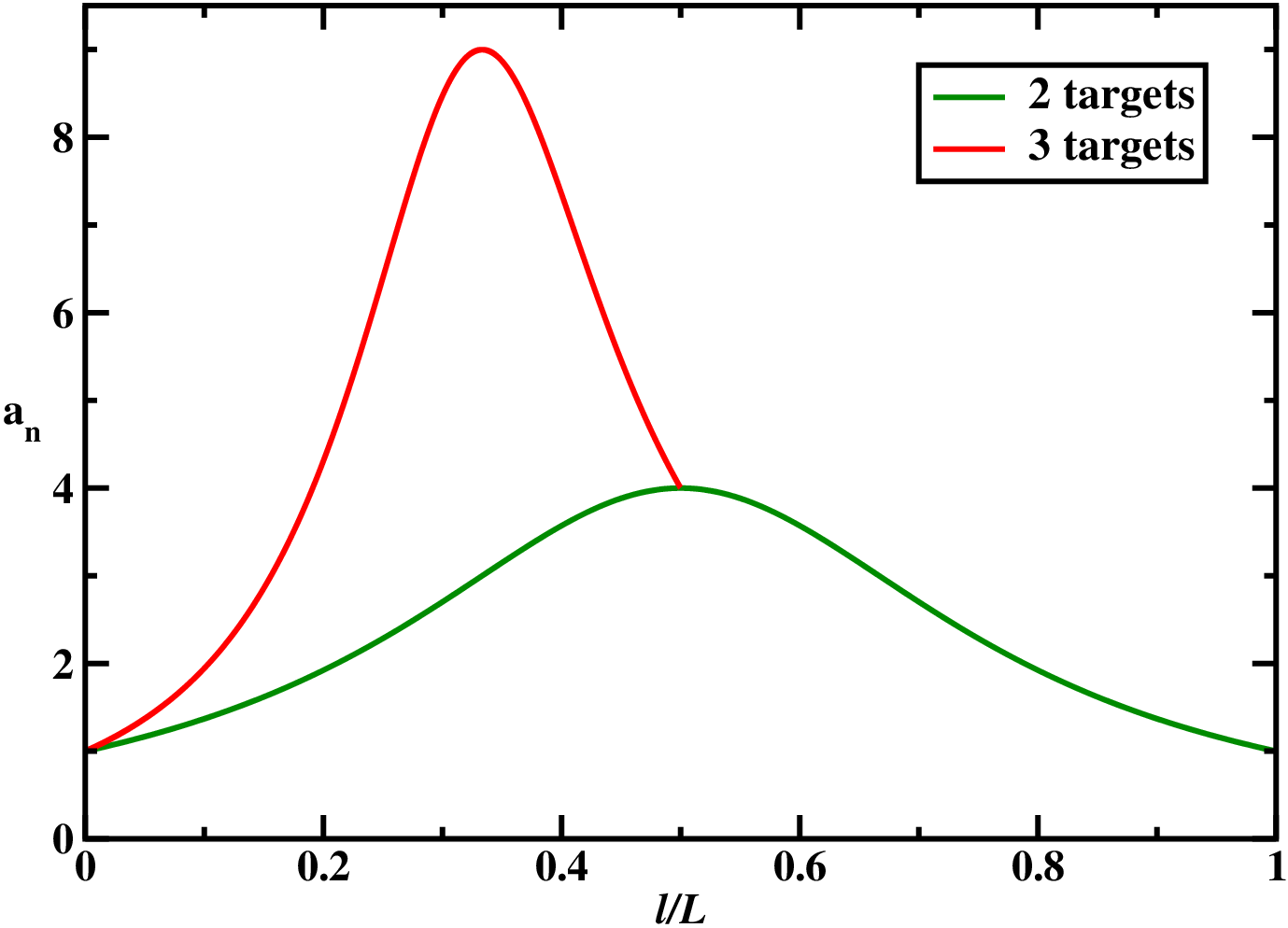}}
    \vskip 1in
    \begin{Large} Figure 6. Lange, Kochugaeva and Kolomeisky \end{Large}
    \caption{}
    \label{fig:6}
  \end{center}
\end{figure}

\newpage

\begin{figure}[h]
  \begin{center}
    \unitlength 1in  
    \resizebox{3.375in}{3.0in}{\includegraphics{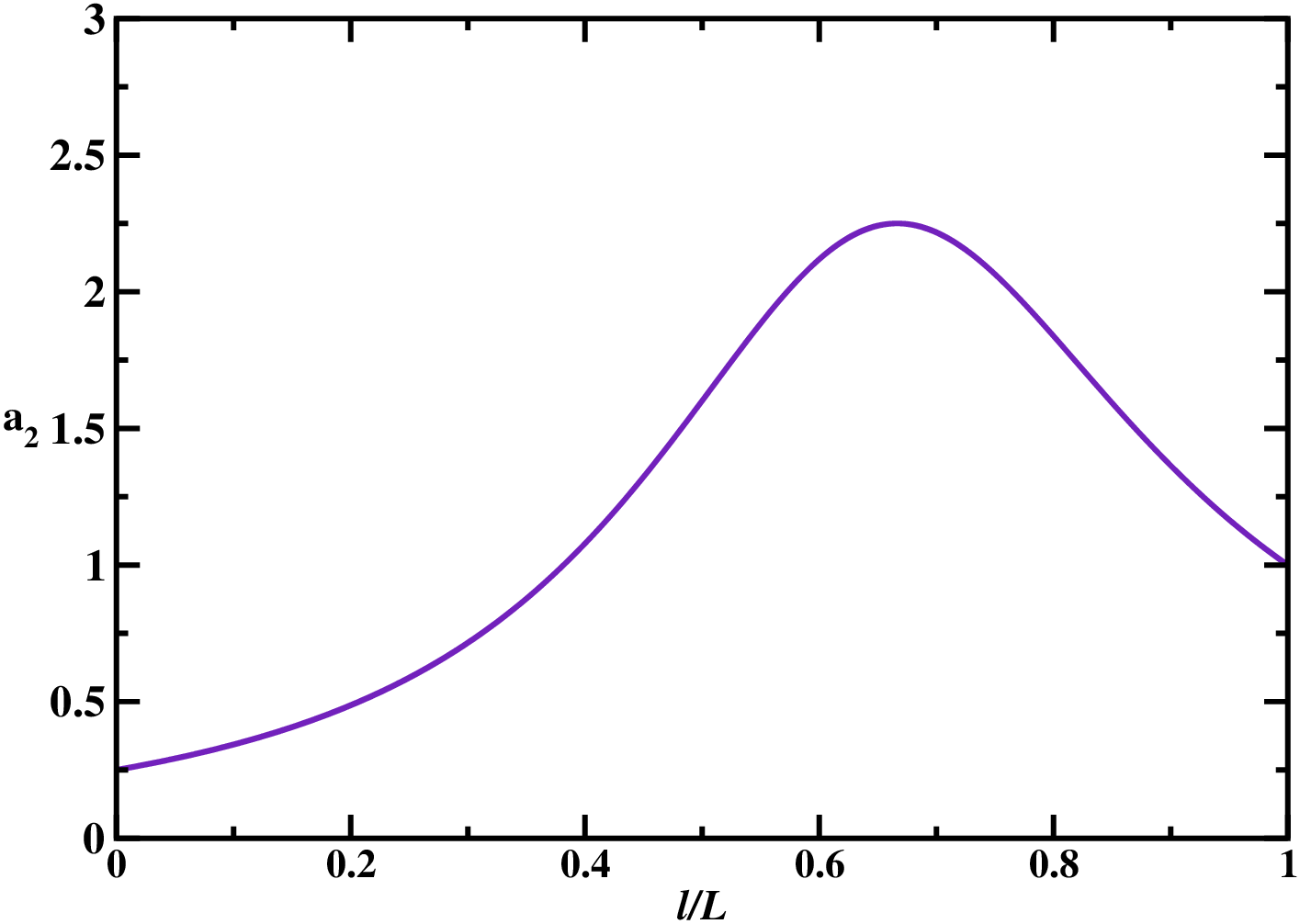}}
    \vskip 1in
    \begin{Large} Figure 7. Lange, Kochugaeva and Kolomeisky \end{Large}
    \caption{}
    \label{fig:7}
  \end{center}
\end{figure}

\newpage

\begin{figure}[h]
  \begin{center}
    \unitlength 1in  
    \resizebox{3.375in}{3.0in}{\includegraphics{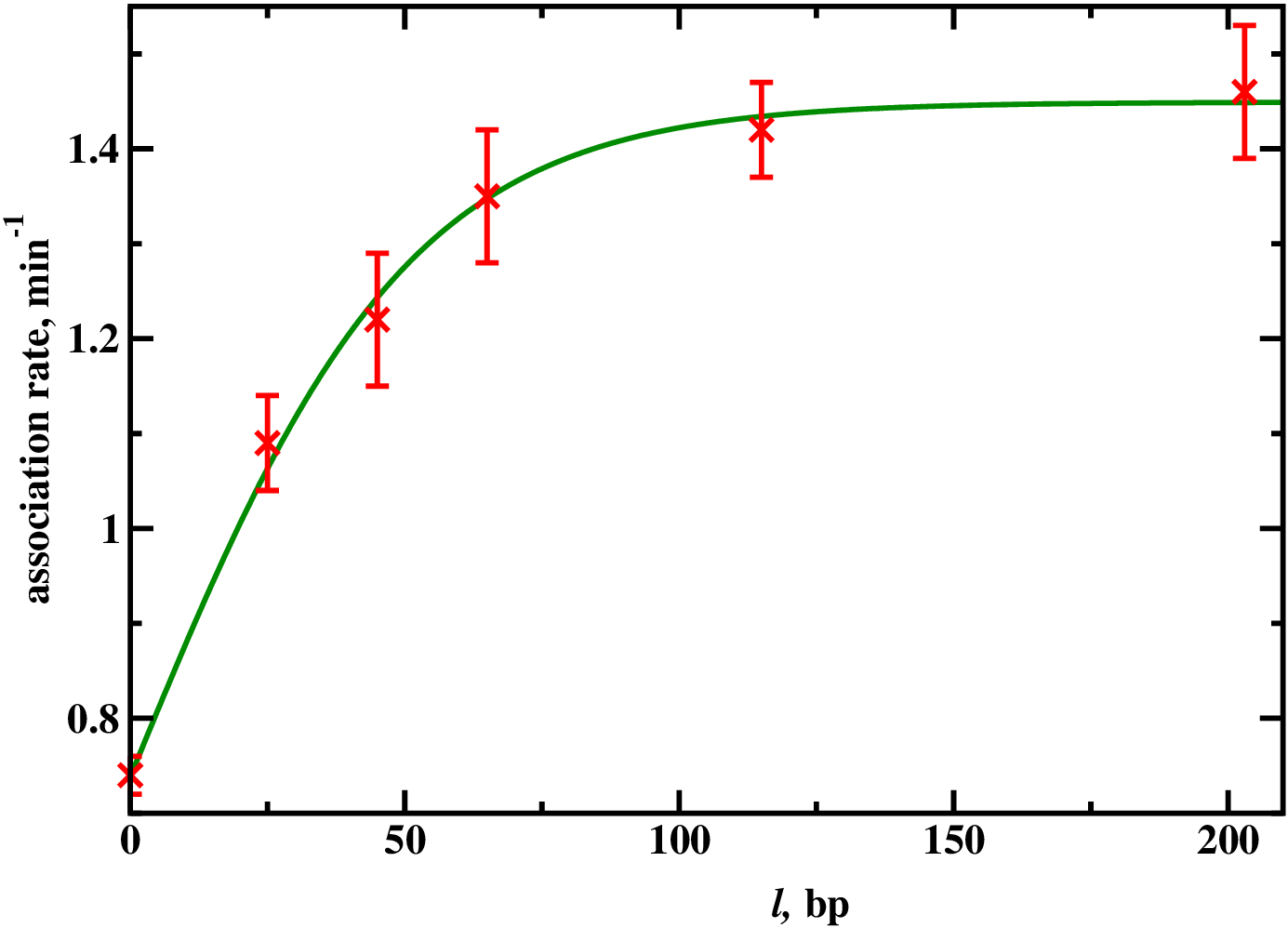}}
    \vskip 1in
    \begin{Large} Figure 8. Lange, Kochugaeva and Kolomeisky \end{Large}
    \caption{}
    \label{fig:8}
  \end{center}
\end{figure}

\newpage

\begin{figure}[h]
  \begin{center}
    \unitlength 1in  
    \resizebox{3.375in}{3.0in}{\includegraphics{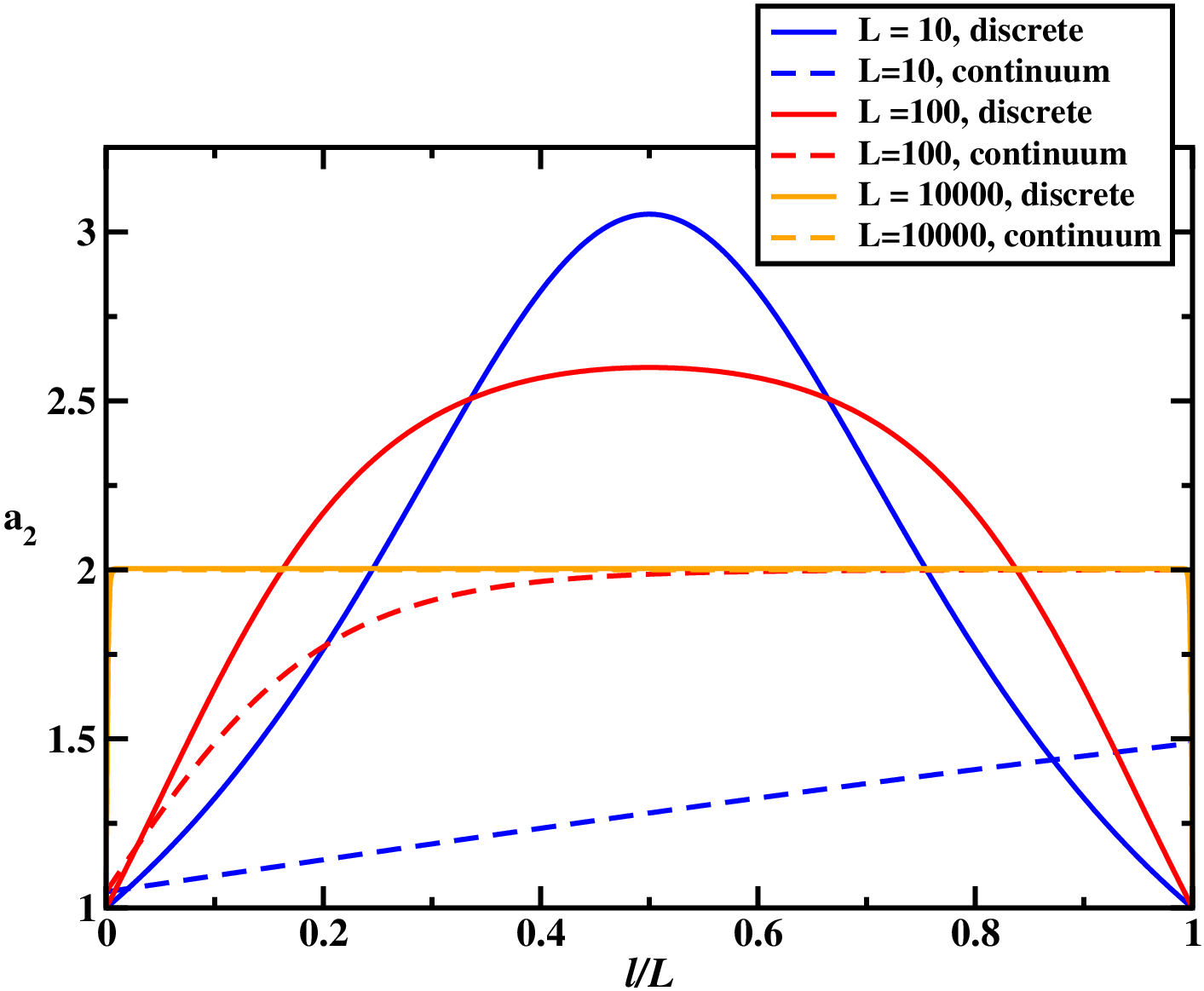}}
    \vskip 1in
    \begin{Large} Figure 9. Lange, Kochugaeva and Kolomeisky \end{Large}
    \caption{}
    \label{fig:9}
  \end{center}
\end{figure}

\newpage

\begin{figure}[h]
  \begin{center}
    \unitlength 1in  
    \resizebox{3.375in}{3.0in}{\includegraphics{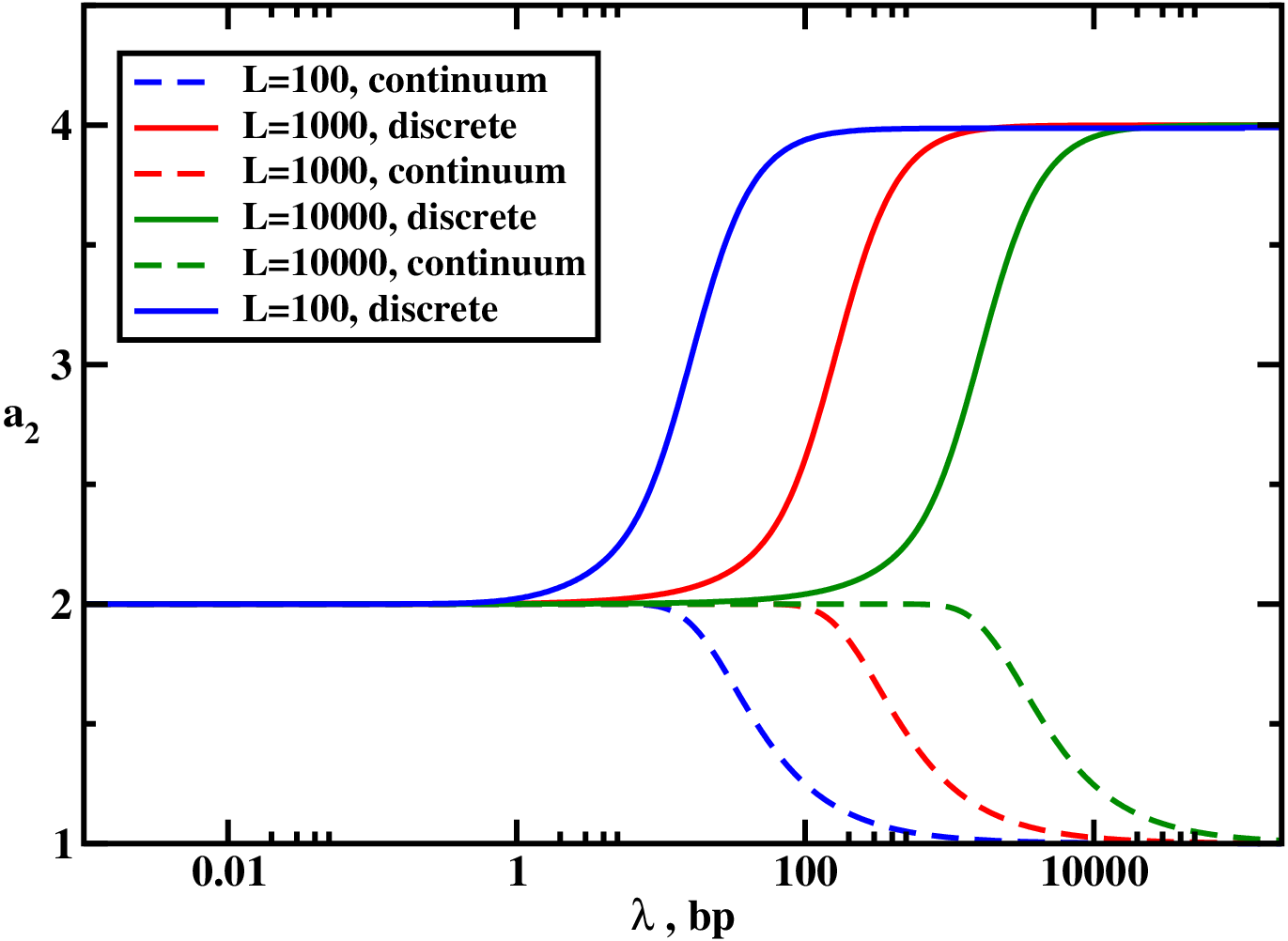}}
    \vskip 1in
    \begin{Large} Figure 10. Lange, Kochugaeva and Kolomeisky \end{Large}
    \caption{}
    \label{fig:10}
  \end{center}
\end{figure}

\end{document}